Large transport critical currents in unsintered MgB$_2$ superconducting tapes

Giovanni Grasso, Andrea Malagoli, Carlo Ferdeghini, Scilla Roncallo, Valeria Braccini and Antonio S. Siri

INFM-Research Unit of Genova, Via Dodecaneso 33, 16146 Genova, Italy

Maria R. Cimberle

CNR/CFSBT, Via Dodecaneso 33, 16146 Genova, Italy

**Abstract**

The Powder-In-Tube process has been employed to fabricate tape-like conductors with a strong metallic sheath and based on the newly discovered $MgB_2$ superconducting phase. Long superconducting tapes have been prepared by packing reacted $MgB_2$ powders inside pure Ag, Cu and Ni tubes which are then cold worked by drawing and rolling. Such tapes have shown transport properties as good as bulk $MgB_2$ samples sintered in high pressure and high temperature conditions. At 4.2 K, the highest critical current density of $10^5$ A/cm$^2$ has been achieved on Nickel-sheathed single-filament conductors. A direct correlation between the sheath tensile strength and the critical current of the unsintered tape has been observed.

The recent discovery of superconductivity at 40 K in the $MgB_2$ compound by Nagamatsu et al. [1] has triggered a great interest of the researchers in applied superconductivity [2]. Immediately after this announcement, the challenge of understanding the grain boundary properties of the $MgB_2$ phase has been addressed by many groups [3-6], with the aim of clarifying whether weak-links would be a limiting factor for the intergrain critical current, as typically occurs in cuprate High Temperature Superconductors.

Bulk $MgB_2$ polycrystalline samples prepared by heat treatment under high pressures have demonstrated that intergrain critical currents at 4.2 K in excess of $10^5$ $A/cm^2$ can be achieved, even with the residual presence of some impurities [7-9]. It is expected that further optimization of the sintering process would raise these $j_c$ values closer to the intrinsic intragrain current density of the $MgB_2$ phase, that has been evaluated to be of the order of $10^6$ $A/cm^2$ still at liquid helium temperature [9]. Nevertheless, by means of magnetization measurements of $MgB_2$ grain agglomerates a few hundred microns in diameter, Bugoslavsky et al. [4] have shown that within these microscopic structures, intergrain and intragrain critical currents can come even closer to each other. Finally, recent results of Sumption et al. [10] have shown that promising transport critical current densities at 4.2 K up to 7.5 x $10^4$ $A/cm^2$ can be achieved within composite Monel-Nb sheathed $MgB_2$ tapes subject to a sintering process at 900°C in Argon atmosphere for several hours.

In the light of these early results and of the constantly growing literature dealing with the preparation of $MgB_2$ conductors [11], the capability of fabricating very dense, Powder-In-Tube (PIT) processed $MgB_2$ tapes with relevant transport currents without applying any sintering process needs to be explored in detail. This opportunity would indeed give much less restrictions to the conductor fabrication route and design, concretely contributing then to a reduction of complexity and cost of the whole tape manufacturing procedure.

Commercially available MgB$_2$ powders ( nominally 98% pure ) manufactured by Alfa-Aesar were loaded into pure metallic tubes ( Ag, Cu, and Ni ) chosen for their common high degree of ductility and malleability, but with different hardness and tensile strength properties. The powders were packed inside tubes of various diameters ( 6 – 8 mm ) and wall thickness ( 1 – 2 mm ) by pressing with a hardened steel piston with a pressure of approximately 250 MPa. The entire loading procedure was carried out in Argon atmosphere, and both tube ends were tightly closed by Tin plugs. An initial packing density of the MgB$_2$ powders of approximately 1.5 g/cm$^3$ was reached by this way. The different tubes were drawn in a number of steps of about 10% of section reduction to a round wire of diameter between 1.5 to 2 mm. These wires were subsequently rolled in following steps of about 10-12% of thickness reduction. The final tape was obtained by rolling the wire to a thickness between 180 and 400 µm, where it presents a width between 3 and 4 mm, while the superconducting fill factor corresponds to about 20 - 30% of the whole conductor volume. A typical transverse cross section of a Cu-sheathed tape of 180 µm in thickness is shown in fig. 1. In this image the MgB$_2$ core of the tape clearly presents a non-homogeneous cross section, that reflects the not fully optimized cold working process.

Furthermore we analyzed the effect of cold working of the above mentioned Cu-sheathed tape on the MgB$_2$ phase by X-ray diffraction measurements. In fig. 2, the θ-2θ scans of the MgB$_2$ phase both before and after cold working are presented. Firstly, in both patterns it is possible to detect the residual presence of a secondary phase, MgO, in addition to the peaks related to the sheath material. However, there is no essential difference between the two X-ray patterns that could be related to a change in texture of the MgB$_2$ grains, which therefore remain substantially isotropic after the cold working. Finally, no sign of evident peak broadening or shifting that could be associated with an increased stress of the MgB$_2$ phase on a microscopic scale has been detected in such Cu-sheathed conductor.

The critical current of the MgB$_2$ tape was measured by the standard four-probe method, with the sample completely immersed in a liquid helium bath. Short pieces ( about 50 mm in length ) of each manufactured conductor ( typical batch length being about 5 meters ) have been measured. Current and voltage leads were directly soldered to the surface of the sheath material with a distance of approximately 10 mm between them. Again, no sintering procedures were employed either during or after the cold working process.

The voltage-current characteristics of the MgB$_2$ tapes were measured at 4.2 K in self field for the different sheath materials we employed. The V-I characteristics presented in fig. 3 were measured on tapes with comparable overall cross section ( 300 µm thick and 3 mm wide ) and superconducting fill factor of about 30%. All the fabricated conductors are actually superconducting in the unsintered state. We have observed that the critical current strongly improves as the sheath material toughness increases, rising from 8 A for the pure Ag-sheathed tape, to 46 A for the pure Cu-sheathed one and to about 80 A for the pure Ni-sheathed tape. The sharpness of the transition from superconducting to normal state also scales with the critical current. In particular, Ni-sheathed tapes show an abrupt transition to the normal state via quench, which is a symptom of a higher intrinsic critical current, as well as of a more problematic current sharing to the resistive matrix, which definitely increases in electrical resistance from pure Ag to pure Ni sheaths.

In the light of these results, a new series of tapes was prepared with a thicker Ni sheath, in order to further increase the density of the $MgB_2$ superconducting filament, and to reduce its transverse cross section at the same time to partly compensate for the stabilization problem. One tape conductor was fabricated by lightly packing ( initial density of 1 $g/cm^3$ ) $MgB_2$ powders into a 12.7 mm x 7.0 mm ( outer x inner diameter ) pure Ni tube that has been drawn in several steps to a wire of 1.8 mm in diameter. A 5 meters long tape was obtained by cold rolling the wire in successive steps to reduce its thickness to about 350 μm as the tape width reached 4 mm. The superconducting fill factor was determined from microscope images of the tape transverse cross section, and represents about 17% of the total conductor cross section.

The critical current of several pieces of this unsintered Ni-sheathed $MgB_2$ tape was measured at 4.2 K, and a selected V-I curve is presented in fig. 3. The 4.2 K critical current values of short pieces of this tape were found to be within the range 240 ± 10 A and, by considering a superconducting cross section of about 0.23 $mm^2$, we can therefore estimate its average transport critical current density to be $10^5$ $A/cm^2$. Again, an abrupt but reproducible transition to the normal state was observed, which is still an indication of the possibility of reaching even larger critical current densities by better stabilizing the superconductor. This is also reflected by the fact that some of the measured conductors burned after unexpected quenches.

In conclusion, we have demonstrated that relevant transport critical current densities can be reproducibly achieved with unsintered Ni-sheathed $MgB_2$ superconducting tapes. This result can be understood by considering that the pressure applied to the superconducting phase during each deformation step of the cold working process is relevant and actually increases with the toughness of the sheath material and with its proportion within the tape. The critical current density of $10^5$ $A/cm^2$ we have reached at present on unsintered Ni-sheathed tapes is comparable with the result achieved by more complex high pressure and high temperature sintering of the same material, but it is still about an order of magnitude lower than the intragrain $MgB_2$ powder current density. However, as the current density level of $10^5$ A/cm$^2$ is considered a common benchmark for magnet applications, it is still necessary to demonstrate the full potential of this new material for high field generation purposes. It is not excluded, however, that the strong pressure applied during the cold working procedure used on the Ni-sheathed tapes could have induced some additional strain in the $MgB_2$ crystal structure, therefore improving their intrinsic superconducting properties. Irreversibility line measurements would definitely help to understand whether this mechanism is actually taking place.

Further enhancement of the tape transport properties is still expected by improving both the tape cold working process and its design which would definitely be facilitated by the fact that the materials eventually chosen to sheath the superconductor are not subjected to any heat treatment. Optimized $MgB_2$ powders with a higher degree of purity and with controlled granularity would also help to improve the cold working process and the density of pinning centers associated with the grain boundaries. On the other hand, the preparation route illustrated in this work would also lead to a minimization of the overall cost of the tape. In fact, expensive barrier materials such Nb or Ta layers that surround the $MgB_2$ phase to prevent its diffusion can be neglected, leaving further room for an improvement of the conductor stabilization.

**Figure Captions:**

Fig. 1: Transverse cross section of a 180 μm thick, 4 mm wide Cu-sheathed $MgB_2$ tape.

Fig. 2: X-ray diffraction patterns of the $MgB_2$ powders before packing and of the filament surface after the cold working process.

Fig. 3: V-I characteristics at 4.2 K of $MgB_2$ tapes with different sheath materials and superconducting fill factor.

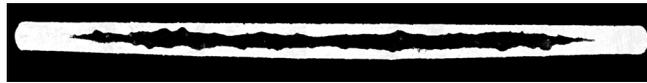

Fig. 1, G. Grasso, APL

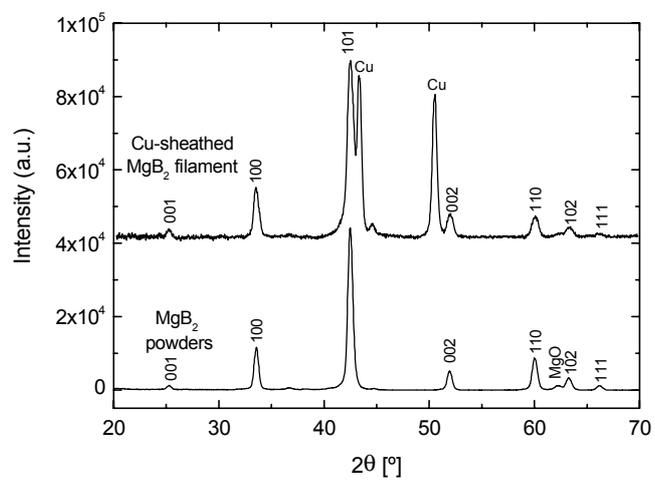

Fig. 2, G. Grasso, APL

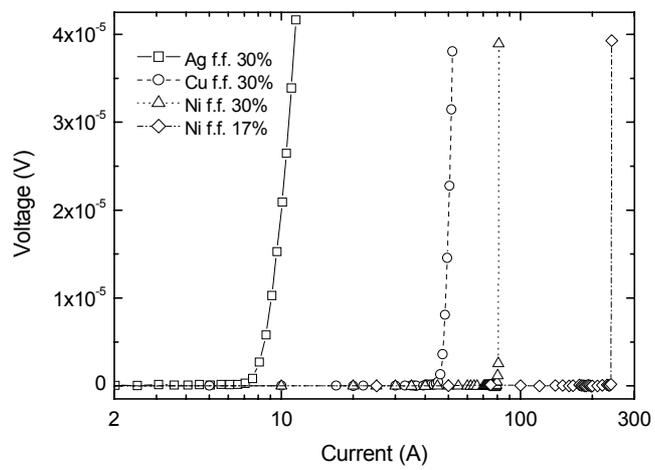

Fig. 3, G. Grasso, APL